# Ultrafast polarization control of zero-bias photocurrent and terahertz emission in hybrid organic perovskites.

Petr A. Obraztsov[1,2], Dmitry Lyashenko[3,4], Pavel A. Chizhov[1], Kuniaki Konishi[5], Natsuki Nemoto[6], Makoto Kuwata-Gonokami[6], Eric Welch[3], Alexander N. Obraztsov[7,2], and Alex Zakhidov[3,4]

[1]A. M. Prokhorov General Physics Institute, Moscow, Russia
[2]Department of Physics and Mathematics, University of Eastern Finland, Joensuu, Finland
[3]MSEC, Texas State University, San Marcos, TX 78666, USA
[4]Department of Physics, Texas State University, San Marcos, TX 78666, USA.
[5]Institute for Photon Science and Technology, The University of Tokyo, Tokyo, Japan
[6]Department of Physics, The University of Tokyo, Tokyo, Japan
[7]Department of Physics, M.V. Lomonosov Moscow State University, Moscow 119991, Russia

(Dated: 26 October 2017)

Methylammonium lead iodide (MAPI) is a benchmark hybrid organic perovskite material, which is used for the low-cost, printed solar cells with over 20% power conversion efficiency. Yet, the nature of light-matter interaction in MAPI as well as the exact physical mechanism behind device operation is currently debated. Here we report room temperature, ultrafast photocurrent and free-space terahertz (THz) emission generation from unbiased MAPI induced by 150 fs light pulses. Polarization dependence of the observed photoresponse is consistent with the Bulk Photovoltaic Effect (BPVE) caused by a combination of injection and shift currents. We believe that this observation of can shed light on low recombination, and long carrier diffusion lengths due to indirect bandgap. Moreover, ballistic by nature shift and injection BPVE photocurrents may enable third generation perovskite solar cells with efficiency that exceed the Shockley–Queisser limit. Our observations also open new venues for perovskite spintronics and tunable THz sources.

The BPVE is a striking quantum phenomenon that is observed in non-centrosymmetric crystals[1]. Unlike conventional p-n junction based PV mechanisms, the BPVE does not require a built-in electric field. There are two proposed mechanisms of the BPVE: injection and shift. The injection (also sometimes called ballistic[2]) currents are associated with hot carriers in a crystal and are caused by the asymmetric distribution of their momenta. Photoexcited non-thermalized carriers lose their energy and descend to the bottom of the band over a free path, which can be hundreds of nanometers[1]. Injection currents constitute the Circular Photogalvanic Effect (CPGE), which depends on the helicity of the incident light. The CPGE has been experimentally observed in quantum wells[3]. It has also been observed in tellurium crystals with significant spin–orbit valence band splitting[4], in topological insulators with spin-splitted surface states[5,6] and 2D transition metal dichalcogenides[7]. In the shift current mechanism, the driving force for carriers is the coherent evolution of electron and hole wave functions due to non-diagonal elements of the density matrix[8]. The shift currents are responsible for the Linear Photogalvanic Effect (LPGE), which depends on the linear polarization.

On a macroscopic level both injection (CPGE) and shift (LPGE) currents can be linked to second-order optical nonlinear effects, as

$$J_\sigma = \beta_{\sigma\mu\nu} e_\mu e_\nu^* E_0^2 \tag{1}$$

where Greek letters subscripts refer to the Cartesian coordinates, $\beta_{\sigma\mu\nu}$ is a third-rank photogalvanic tensor, $\vec{e}$ is a unit vector of light wave polarization and $E_0^2$ is light intensity. For both mechanisms, the current dynamics are expected to be on a subpicosecond time scale; thus the BPVE currents are often accompanied by free-space THz emission upon femtosecond optical excitation[9]. Moreover, hot carrier harvesting enables generation of photovoltage that can greatly exceed the band gap[10]. Therefore, solar cell devices based on the BPVE can theoretically have a power conversion efficiency (PCE) that exceeds Shockley–Queisser limit[11] for excitonic PV cells[12].

MAPI has a tetragonal $ABX_3$ perovskite structure at room temperature with a polar molecular cation $CH_3NH_3^+$ at each A-site[13–19]. It was recently predicted that it is possible to observe the BPVE for the polar configurations of the MAPI crystal structure[20,21]. The absence of inversion symmetry, together with the strong spin-orbit coupling in the heavy Pb element, leads to the Rashba-type spin splitting with a preferred spin orientation perpendicular to both the momentum and the inversion symmetry breaking direction[22]. Density Functional Theory (DFT

) calculations show that such spin texture can lead to the CPGE, where, due to spin-momentum locking and angular momentum selection rules, optical transitions between conduction and valence energy states respond differently to light with right and left circular polarization[23].

Here, for the first time, we demonstrate the generation of polarization sensitive ultrafast photocurrents in unbiased MAPI at room temperature by performing photocurrent measurements and free-space terahertz experiments. By measuring the amplitude and time-domain waveforms of the photocurrents under excitation with different optical polarizations, incidence angle, and wavelengths, we extract the contributions of injection photocurrents (sensitive to helicity) and shift photocurrents (sensitive to linear polarization) from the overall photocurrent response. The simultaneous observation of both "circular" and "linear" photocurrents is direct evidence of a strong BPVE in MAPI, which provides a fundamentally new route for third-generation photovoltaic solar energy conversion.

The optical scheme of the experimental setup is shown in the Fig. 1a. For these experiments, 350nm thick MAPI film was deposited on a soda lime glass substrate with pre-patterned 120nm thick indium tin oxide (ITO) electrodes (see *Methods*). MAPI films were optically excited either with a single beam of continuous wave (CW) laser diodes operating at 532 nm and 650 nm, or radiation of fundamental (780 nm) and second harmonics (390 nm) of femtosecond pulsed lasers (see *Methods*). Photoinduced currents were measured as a voltage drop, $U \propto J$, across a 50 Ω load resistor and recorded in unbiased samples with a storage oscilloscope. By changing the orientation of MAPI sample, we were able to measure longitudinal ($J_x$) and transverse ($J_y$) currents with respect to the light incidence plane components of the induced photocurrent. Polarization dependencies of the induced $J_x$ and $J_y$ were identified by measuring peak-to-peak amplitude of the corresponding photocurrent waveform while rotating quarter ($\lambda/4$) waveplate positioned before the sample by an angle $\alpha$, which varied the state of laser polarization with a 90° and 180° period respectively. Fig. 1b shows waveforms of $J_y$ components obtained at 45° and -45° incidence angles are shown. Figs. 1c,d show the dependence of respective photocurrents $J_x$ and $J_y$ excited with femtsceond pulses at 390 nm on the $\lambda/4$ wavevplate angle. This dependence is the most informative to start our analysis of the photocurrent generation in MAPI as it should contain contributions from all possible mechanisms. The presented data shows that both $J_x$ and $J_y$ components of the photocurrent induced in MAPI by obliquely incident light exhibit strong polarization dependence and change their sign upon reversing the direction of incidence from 45° to -45°.

The overall dependence of *J* on *α* demonstrate a complex behavior, which is well described by an oscillating function comprised of four components:

$$J(\alpha) = D + C \cdot \sin(2\alpha) + L1 \cdot \cos(4\alpha) + L2 \cdot \sin(4\alpha), \qquad (2)$$

where *D* is a polarization independent offset, ascribed to laser-heating gradients or the photo-Dember effect in the sample (we confirm the origin of this offset by sweeping the laser spot across two electrodes with a fixed polarization; see Fig. S2). This offset is presented in all the data obtained for both $J_x$ and $J_y$ components. To prevent the temperature gradient between contacts necessary for the thermoelectric effect and to minimize the offset *D* in our measurements, the laser spot was always centered between the contacts. However, due to the finite size of the laser spot on the sample, some small polarization independent contribution offset was always observed. The coefficient *D* deviates from measurement to measurement as we used different laser sources and samples. The coefficient *C* describes the strength of circular (helicity-dependent) contribution to the photocurrent because rotating the $\lambda/4$ waveplate varies the light polarization between left- (σ-) and right- (σ+) circular with the functional form *sin(2α)*. The coefficients *L1* and *L2* in (1) parameterize the helicity-independent photocurrents caused by rotation of linear polarization with the functional forms *cos(4α)* and *sin(4α)* respectively. For the *α* angles of 45° and 135° (purely circular polarization) the linear contribution vanishes.

The principal observation made from polarization dependent measurements is that oblique incidence excitation of MAPI sample with right and left circular polarized light leads to generation of transverse ($J_y$) photocurrents flowing along opposite directions while the longitudinal ($J_x$) photocurrents do not "feel" the helicity of light and depend only on rotation of linear polarization. The amplitudes of polarization sensitive $J_x$ and $J_y$ photocurrents linearly depend on the radiation intensity. We should also note here that the helicity dependence completely vanishes at normal incidence while minor modulation of $J_x$ and $J_y$ amplitudes with change of linear polarization angle persists even under normally incident light (Fig. S3).

The overall behavior of the observed photocurrents falls under the class of second-order nonlinear effects, which includes circular-, and linear- photogalvanic effects (also known as injection and shift currents or bulk photovoltaic effect[8]) and the photon-drag effect (also known as the AC Hall effect[25] or light-induced drift currents[26]). The fundamentally distinct feature of this class of effects comparing to conventional photovoltaic optical effects where charge carrier's separation (photocurrent) is only possible when internal or external fields are applied, is the possibility to observe intrinsic and instantaneous photocurrent response of excited carriers with no external influence on the carriers. In general, the second order optical response is allowed in the material if either (i) the spatial inversion symmetry of the material is broken or (ii) second-order conductivities change their sign at spatial inversion [i.e. there is linear coupling between the current and photon wavevector (momentum)[8]]. The second case is related to the momentum transfer from a photon to the excited carrier. While both circular and linear photogalvanic effects are possible only in the systems that lack inversion symmetry, the photocurrent proportional to photon momentum does not require central symmetry and, it is therefore allowed in any medium. Similar to the photogalvanic effect, the photon drag effect in principle takes place in response to both linearly and circularly polarized radiation. Thus, these two phenomena demonstrate indistinguishable dependence on light polarization. However, there is a significant difference between effects in terms of wavevectors. In particular, the photogalvanic effect is an even function of the wavevector (photoresponse does not change sign when $k$ is replaced with $-k$) while photon-drag is odd (i.e. photoresponse do change sign). Therefore, the photocurrents associated with the photogalvanic and photon-drag effects demonstrate distinct behavior and can be distinguished in the experiment by reversing the excitation beam direction. In the experiment the reverse of the wavevector direction can be realized via applying front and back illumination of the sample[27] (Note: There is no difference between back and front excitation in terms of the sample geometry. As shown on Fig. S4 the MAPI film is covered with encapsulation glass of the same quality as substrate). To elucidate the particular mechanisms of the polarization sensitive photocurrents generated in MAPI samples, we further focus on analyzing the transverse component $J_y$ under the back and the front excitation. The experimental dependencies of photocurrent amplitude on the quarter-waveplate angle obtained under front and back illumination of the MAPI sample incident at 45° with a femtosecond laser at 390 nm are summarized in Fig. 2. The data is fitted with Equation (2) and the percentile contributions of different types of photocurrents extracted from fitting are shown on a bar plot on the right side of the figure. Strikingly, at first glance, the photocurrents excited from the front and the back sides of the sample differently, depend on the polarization state, while the average amplitude of the photocurrent signal remains constant. Namely, the contribution of photocurrents depending on linear polarization ($L1$ and $L2$ coefficients in equation (2)) decreases in the case of back side excitation. The amplitude value of the circular photocurrent ($C$) in contrast is larger in this case. To exclude any influence of the pulsed nature of the light on the amplitude and the sign of $J_y$ photocurrents excited from the front and back sides, we performed independent measurements with a CW laser. Excitation of MAPI sample with a CW laser leads to a DC photocurrent response, which can be detected, via occurrence of a constant positive or negative (depending on sample orientation and incidence polarization) offset on the oscilloscope. The photocurrent traces taken with CW lasers operating at 650 nm and 532 nm under excitation from both sides of the MAPI sample are shown in Fig. 2. The data taken with CW lasers is in agreement with femtosecond data and qualitatively confirms the decrease of linear $L1$ and $L2$ contributions and increase of circular current $C$ upon changing the direction of excitation. A decrease in the linear contribution with the change in the direction of propagation of light evidences the assumption that there are two types of currents responsible for the dependence on linear polarization. Indeed, when exciting the sample from the front side, the dependence on linear polarization is due to combination of parallel linear photogalvanic (shift) and photon drag currents. When changing the direction of excitation, these currents become antiparallel: photon-drag current changes sign while the direction of the shift current remains the same. If one assumes that the amplitudes of shift and photon-drag currents are the same, in the case of back excitation, they should compensate each other, and in turn, the linear contribution to photoresponse would become zero. In reality, the amplitudes and temporal dynamics of the shift and photon-drag currents are not equal and might also depend on excitation wavelength. This qualitative picture is in agreement with the experimental data presented in Fig. 3 where complete compensation of the linear component is never achieved.

At the excitation with femtosecond pulses, the duration (width) of the induced AC response obtained from the MAPI sample is determined by the registration system bandwidth rather than by the relaxation time of the excited carriers, while its magnitude represents the time-integrated current. As the photocurrent is triggered by a sub-picosecond-long laser pulse, and the current dynamics are also expected to be on a subpicosecond time scale, the time-varying current $J$ can result in the emission of the THz wave in proportion to $\partial J/\partial t$. As these photoexcitations occur in the unbiased sample, the transient photocurrents $J_x$ and $J_y$ or the resultant free-space emission of $p$- or $s$- polarized THz radiation respectively provides information about the internal bias near the sample surface or the relevant electrodynamics parameters. Due to the limited time resolution in photocurrent experiments, it is difficult to conclude whether charge current is directly induced by light illumination, without information on the transient response of the current, or indirectly by some other effects e.g. photo-Dember or thermo-effect. In order to detect the instantaneous polarization-sensitive photocurrent response upon excitation with femtosecond light pulses in the MAPI samples, we employed time-resolved THz emission spectroscopy. Typical transient waveforms of the emitted THz radiation detected with electro-optic sampling at the far field under oblique excitation with linear and circular photon polarizations are shown in Fig. 3a. The emitted THz pulses have a subpicosecond duration and consist of a nearly single cycle oscillation. The shape of the THz waveform remains essentially constant for all incident polarizations and fluences. The THz frequency spectra obtained by Fourier transformation of the corresponding transient waveform are shown in Fig. 3(a) and (b). The frequency of the maximum THz amplitude is about 0.7 THz. Theoretically, the time evolution of shift currents differs from that of injection and photon-drag currents[28]. The shift current directly follows the envelope of the optical pulse intensity, whereas only the onset of other currents follows this envelope; these currents then decay with the momentum scattering time (relaxation time). However, in practice, due to very fast momentum scattering of carriers, the current dynamics are expected to be the same on a time scale of several 100 fs, which is in agreement with virtually identical THz time-domain waveforms obtained under excitation with linear and circular polarization (see Fig.3a).

To additionally confirm that the emitted THz radiation is due to the same mechanism as the photocurrents presented in Figs. 2 we varied the polarization of the pump (second-harmonic) beam using a quarter-waveplate. In Fig. 3(c), the y-component of the generated THz electric field is plotted as a function of the polarization state of the incident pump light. The measured polarization dependence resembles those obtained with contact measurements and is well fitted with equation (2). The results of fitting are shown in the inset of Fig. 4c.

Since all our optical measurements were done with the same film that is used for PV cells while at room temperature, we believe that our findings can also help to understand the photophysics of the perovskite solar cell devices, and ultimately engineer devices with record efficiencies that would challenge the S-Q limit. First, from observation of the CPGE we conclude that MAPI films at room temperature lack inversion symmetry. We also note that inversion symmetry breaking in our experiments can be caused by light-induced polarons, which lead to collective distortion of the crystal lattice[29,30]. In fact the presence or absence of polarons in MAPI films might explain controversial reports on MAPI polarity[31,32].

Second, our experimental data confirms theoretically predicted Rashba spin-splitting of bands and a resulting indirect band gap[22], which can explain low recombination rates and high carrier diffusion length observed for MAPI[33,34,35,36]. This mechanism was recently questioned by T. Etienne et. al. who performed DFT calculation for dynamical MAPI systems and concluded that small splitting magnitude (<10 mEV) might have a marginal effect on the reduction of the carrier recombination (i.e. less than an order of magnitude)[37]. Our DFT+U calculations (with Hubbard correction and spin orbit coupling) of the MAPI band structure in the presence of hole polarons, however, yield much higher splitting values of 30 meV (Fig. S6). Here it is worth noting that Rashba spin splitting is caused by the local arrangements of the atoms in the unit cell rather than by the average, long-range symmetry of the crystal[38].

Finally, THz emission from MAPI films implies ultrafast photocurrents with response times comparable to the carrier thermalization times (in the order of few ps[33]). Thus, high $V_{OC}$ close to the bandgap and even slightly above it can be obtained for perovskite PV cells with appropriate transport layers. Moreover, as it was recently pointed out by Spanier et al.[12] and Tan[20] photoinduced injection and shift currents can be used to build PV device architectures with power conversion efficiencies that go beyond S-Q limit.

In conclusion, we present experimental evidences for optically stimulated ultrafast injection and shift currents in MAPI films at room temperature. Our measurements confirm the polar nature of MAPI films and an indirect bandgap due to Rashba splitting at room temperature. This result can help to understand device photophysics of highly efficient perovskite solar cells and in particular recombination suppression and long carrier diffusion lengths. Moreover, observed ultrafast injection shift currents could enable third generation perovskite PV cells with efficiencies that break S-Q limit. Our findings also open new venues for perovskite spintronics, ultrafast photodetectors, and tunable THz emitters.

## Methods:

Sample fabrication:
Following Lee et al we have adopted the perovskite deposition method using a $CH_3NH_3I+PbCl_2$ (3:1) solution precursor[14]. Perovskite precursors and a dimethylformamide (DMF) solvent were purchased from Sigma Aldrich and used as received. Patterned ITO-glass substrates were cleaned via 20 minutes of ultrasonic bath in 5 wt% of Deconex OP121 detergent in a DI water solution. After the ultrasonic bath, substrates were rinsed with DI water, dried at 200 $^{o}$C for 60 minutes on a hot plate and treated with oxygen plasma (Harrick Plasma, Pdc-32G) for 5 minutes. Perovskite film of c.a. 350nm was formed from a precursor solution ($CH_3NH_3I$ and $PbCl_2$ at a 3:1 molar ratio dissolved in anhydrous DMF) by spin casting at 4000 rpm followed by 120min annealing at 90 $^{o}$C. The film was encapsulated with a top recess glass slide with edges being sealed by the NOA68 UV epoxy. This method results in formation of MAPI films with uniform coverage and long-range crystalline domains. Fabricated with this deposition method PV cells (ITO/PEDOT:PSS/MAPI/$C_{60}$/BCP/Al) show reproducible PCE of 12±2% under 1.5 AM, 100 mW/cm$^2$ illumination, which is comparable with state of the art PV cells of a similar architecture and film deposition method[24] (Fig. S1). X-Ray Diffraction (XRD) and scanning electron microscopy (SEM) images of the microcrystalline MAPI film are both consistent with typical data reported for multicrystalline perovskite films that are highly oriented with the $\alpha$ axis (Fig. S1).

Photocurrent response measurements:
In order to induce photocurrents in a MAPI sample we employed a Ti:Sapphire amplified system (Coherent Elite Pro) delivering 150 fs pulses with 1 kHz repetition rate at 780 nm (1.59 eV) fundamental wavelength and output pulse energy of 2.6 mJ. In the experiments we used the second harmonic fundamental wavelength centered at 390 nm (3.18 eV) generated in a beta-barium-borate (BBO) crystal. The particular pump energies on the sample where 500 µJ and 30 µJ for second harmonics. The used pump fluencies are well below the damage threshold of our samples and did not result in any damages and/or artifacts in our measurements. To reveal the dependence of photocurrent response on excitation laser wavelength and to confirm the possibility of photocurrent generation under excitation with CW radiation, we also employe two different low-power diode lasers operating at 532 nm (2.3 eV) and 650 nm (1.9 eV) with less than 1mW average power. The induced photocurrents ($J$) were measured at room temperature across unbiased MAPI samples by the voltage drop on the 50Ω input impedance of a 600 MHz digital oscilloscope connected to ITO electrodes under short circuit conditions, or an A.C. lock-in amplifier in the case of excitation with a CW laser. The MAPI sample was excited in two different ways, depending on the experiment. In one case, the pump beam excites the sample from the front side through the cap-glass slide as shown in (Fig.1). The MAPI sample can also be excited from the back side through the glass substrate. By changing the orientation of the MAPI sample we were able to measure longitudinal ($J_x$) and transverse ($J_y$) components with respect to the light incidence plane components of the induced photocurrent. In all measurements the pump beam was directed to the MAPI surface without any focusing. The incidence angle $\Theta$ was varied with a rotation stage between -45$^{o}$ and 45$^{o}$ while the excitation beam spot size on the sample was controlled with an iris aperture and varied from 200 µm to 0.5 cm depending on the experiment. Polarization dependencies of the induced $J_x$ and $J_y$ component of photocurrents were identified by measuring peak-to-peak amplitudes of the corresponding photocurrent waveform while rotating half ($\lambda/2$) or quarter ($\lambda/4$) waveplates positioned before the sample by an angle $\alpha$, which varied the state of laser polarization with a 90$^{o}$ and 180$^{o}$ period respectively.

Terahertz emission measurements:

The terahertz emission experiment was performed using a standard THz time-domain spectroscopy setup based on electro-optic sampling using a (111) oriented 450-µm-thick GaP crystal. In this experiment, we employed as a light source, a femtosecond diode-pumped Yb:KGW regenerative amplifier (PHAROS-SP1.5 mJ, Light Conversion, Ltd.) with a central wavelength of 1028 nm, repetition rate of 14.0 kHz, average power of 5.5 W, and pulse duration of 200 fs. The MAPI sample was excited with second harmonic radiation centered at 514 nm generated in a BBO crystal. The emitted THz radiation was collected with a system of golden parabolic mirrors in the direction of laser beam reflection as shown on (Fig. S5). The orientation of a GaP electro-optic (EO) crystal was chosen in such a way that only *s*-polarized THz radiation associated with transversal photocurrent $J_y$ is detected. The delayed probe beam was used for time-equivalent sampling of the THz time-domain waveforms employing a typical balanced detection scheme. The polarization of the excitation second-harmonic beam was controlled with a quarter-waveplate. All the measurements were performed at room temperature in ambient atmosphere.

# References:

1. Sturman, B. & Fridkin, V. *The Photovoltaic and Photorefractive Effects in Noncentrosymmetric Materials*. (CRC Press, 1992).
2. Belinicher, V. I. & Sturman, B. I. The relation between shift and ballistic currents in the theory of photogalvanic effect. *Ferroelectrics* **83,** 29–34 (1988).
3. Glazov, M. M. & Ganichev, S. D. High frequency electric field induced nonlinear effects in graphene. *Phys. Rep.* **535,** 101–138 (2014).
4. Shalygin, V. a., Moldavskaya, M. D., Danilov, S. N., Farbshtein, I. I. & Golub, L. E. Circular photon drag effect in bulk tellurium. *Phys. Rev. B - Condens. Matter Mater. Phys.* **93,** 1–8 (2016).
5. McIver, J. W., Hsieh, D., Steinberg, H., Jarillo-Herrero, P. & Gedik, N. Control over topological insulator photocurrents with light polarization. *Nat. Nanotechnol.* **7,** 96–100 (2011).
6. Braun, L. *et al.* Ultrafast photocurrents at the surface of the three-dimensional topological insulator B2Se3. *Nat. Commun.* **7:13259,** 1–9 (2015).
7. Yuan, H. *et al.* Generation and electric control of spin–valley-coupled circular photogalvanic current in WSe2. *Nat. Nanotechnol.* **9,** 851–857 (2014).
8. Von Baltz, R. & Kraut, W. Theory of the bulk photovoltaic effect in pure crystals. *Phys. Rev. B* **23,** 5590–5596 (1981).
9. Priyadarshi, S. *et al.* Terahertz spectroscopy of shift currents resulting from asymmetric (110)-oriented GaAs/AlGaAs quantum wells. *Appl. Phys. Lett.* **95,** 10–13 (2009).
10. Yang, S. Y. *et al.* Above-bandgap voltages from ferroelectric photovoltaic devices. *Nat. Nanotechnol.* **5,** 143–147 (2010).
11. Shockley, W. & Queisser, H. J. Detailed Balance Limit of Efficiency of p-n Junction Solar Cells. *J. Appl. Phys.* **32,** 510 (1961).
12. Spanier, J. E. *et al.* Power conversion efficiency exceeding the Shockley-Queisser limit in a ferroelectric insulator. *Nat. Photonics* **10,** 611–616 (2016).
13. Im, J.-H., Lee, C.-R., Lee, J.-W., Park, S.-W. & Park, N.-G. 6.5% efficient perovskite quantum-dot-sensitized solar cell. *Nanoscale* **3,** 4088–4093 (2011).
14. Lee, M. M., Teuscher, J., Miyasaka, T., Murakami, T. N. & Snaith, H. J. Efficient Hybrid Solar Cells Based on Meso-Superstructured Organometal Halide Perovskites. *Science* **338,** 643–647 (2012).
15. Etgar, L. *et al.* Mesoscopic CH3NH3PbI3/TiO2 heterojunction solar cells. *J. Am. Chem. Soc.* **134,** 17396–17399 (2012).
16. Crossland, E. J. W. *et al.* Mesoporous TiO2 single crystals delivering enhanced mobility and optoelectronic device performance. *Nature* **495,** 215–9 (2013).
17. Heo, J. H. *et al.* Efficient inorganic–organic hybrid heterojunction solar cells containing perovskite compound and polymeric hole conductors. *Nat. Photonics* **7,** 486–491 (2013).
18. Yang, W. S. *et al.* High-performance photovoltaic perovskite layers fabricated through intramolecular exchange. *Science* **348,** 1234–1237 (2015).
19. Manspeaker, C., Venkatesan, S., Zakhidov, A. & Martirosyan, K. S. Role of interface in stability of perovskite solar cells. *Curr. Opin. Chem. Eng.* **15,** 1–7 (2017).

**20.** Zheng, F. et al. First-Principles Calculation of the Bulk Photovoltaic Effect in $CH_3NH_3PbI_3$ and $CH_3NH_3PbI_{3-x}Cl_x$. *J. Phys. Chem. Lett.*, **fDF**), 31-37 (2015).

Tan, L. Z. *et al.* Shift current bulk photovoltaic effect in polar materials—hybrid and oxide perovskites and beyond. *npj Comput. Mater.* **2,** 16026 (2016).
Leppert et al., J. Phys. Chem. Lett. **7**, 3683 (2016).
21. Li, J. & Haney, P. M. Circular photogalvanic effect in organometal halide perovskite CH3NH3PbI3. *Appl. Phys. Lett.* **109,** 193903 (2016).
22. Zheng, F., Tan, L. Z., Liu, S. & Rappe, A. M. Rashba spin-orbit coupling enhanced carrier lifetime in CH3NH3PbI3. *Nano Lett.* **15,** 7794–7800 (2015).
23. Li, J. & Haney, P. M. Optical spintronics in organic-inorganic perovskite photovoltaics. *Phys. Rev. B - Condens. Matter Mater. Phys.* **93,** 1–9 (2016).
24. You, J. *et al.* Perovskite Solar Cells with High Efficiency and Flexibility. 1674–1680 (2014).
25. Valov, P. M. *et al.* Dragging of electrons by photons in intraband absorption of light by free carriers in semiconductors. *Sov. Phys. JETP* **32,** 1038–1041 (1971).
26. Shalaev, V. M., Douketis, C., Stuckless, J. T. & Moskovits, M. Light-induced kinetic effects in solids. *Phys. Rev. B* **53,** 11388–11402 (1996).
27. Obraztsov, P. a. *et al.* Photon-drag-induced terahertz emission from graphene. *Phys. Rev. B - Condens. Matter Mater. Phys.* **90,** 1–5 (2014).
28. Laman, N., Bieler, M. & van Driel, H. M. Ultrafast shift and injection currents observed in wurtzite semiconductors via emitted terahertz radiation. *J. Appl. Phys.* **98,** 103507 (2005).
29. Welch, E., Scolfaro, L. & Zakhidov, A. Density Functional Theory + U modeling of polarons in organohalide lead perovskites. *AIP Adv.* **6,** 125037 (2016).
30. Neukirch, A. J. *et al.* Polaron stabilization by cooperative lattice distortion and cation rotations in hybrid perovskite materials. *Nano Lett.* **16,** 3809–3816 (2016).
31. Sharada, G. *et al.* Is CH3NH3PbI3 Polar? *J. Phys. Chem. Lett.* **7,** 2412–2419 (2016).
32. Kutes, Y. *et al.* Direct Observation of Ferroelectric Domains in Solution-Processed CH3NH3PbI3 Perovskite Thin Films. *J. Phys. Chem. Lett.* **5,** 3335–3339 (2014).
33. Xing, G. *et al.* Long-Range Balanced Electron- and Hole-Transport Lengths in Organic-Inorganic CH3NH3PbI3. *Science* **342,** 344–347 (2013).
34. Dong, Q. *et al.* Electron-hole diffusion lengths >175 um in solution grown CH3NH3PbI3 single crystals. *Science* **347,** 967–970 (2015).
35. Stranks, S. D. *et al.* Electron-hole diffusion lengths exceeding 1 micrometer in an organometal trihalide perovskite absorber. *Science* **342,** 341–344 (2013).
36. Chen, Y. *et al.* Extended carrier lifetimes and diffusion in hybrid perovskites revealed by Hall effect and photoconductivity measurements. *Nat. Commun.* **7,** 12253 (2016).
37. Etienne, T., Mosconi, E. & De Angelis, F. Dynamical Origin of the Rashba Effect in Organohalide Lead Perovskites: A Key to Suppressed Carrier Recombination in Perovskite Solar Cells? *J. Phys. Chem. Lett.* **7,** 1638–1645 (2016).
38. Zhang, X., Liu, Q., Luo, J.-W., Freeman, A. J. & Zunger, A. Hidden spin polarization in inversion-symmetric bulk crystals. *Nat. Phys.* **10,** 387–393 (2014).

## Acknowledgements

This work was partially supported by the Russian Science Foundation Grant # 17-72-10303, the Academy of Finland Grant #299059, the ACS Petroleum Research Fund Grant #56095-UNI6 (Program Manager Askar Fahr) and the U.S. DoD Contract W911NF-16-1-0518 (Program officer Paul Armistead). This work is also supported by JSPS international joint research program.

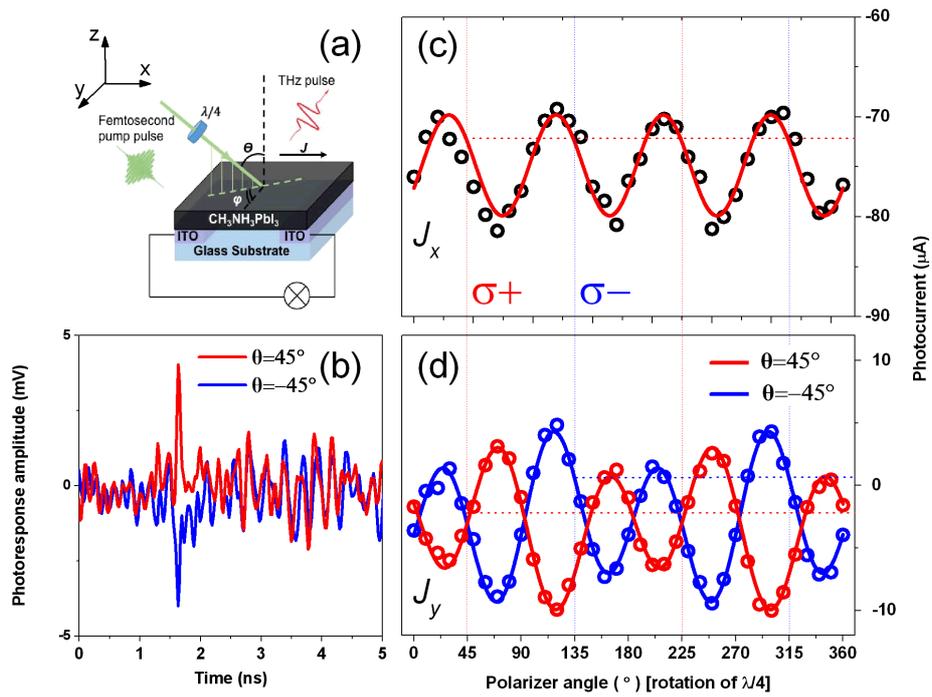

Figure 1 | Polarization dependent photocurrent in unbiased MAPI sample at room temperature. Configuration (a) and temporal profiles of photocurrent response measured (b) at different incidence angles (θ=45°: red color; θ=-45°: blue color) in the transversal geometry of the experiment. (c) and (d) Excitatation beam polarization dependence of correspondingly longitudinal ($J_x$) and tranversal ($J_y$) components of photocurrent response measured at 45° incidence. The excitation beam polarization state is controlled by rotation of the quarter wave-plate with a step of 5 degrees. 45° and 225° angles correspond to left circular polarization; 135° and 315° correspond to right circular polarization. The open circles are the experimental data. The fitting data with function $J=J_0+C\cdot\sin(2\alpha)+L1\cdot\cos(4\alpha)+L2\cdot\sin(4\alpha)$ are shown with solid lines.

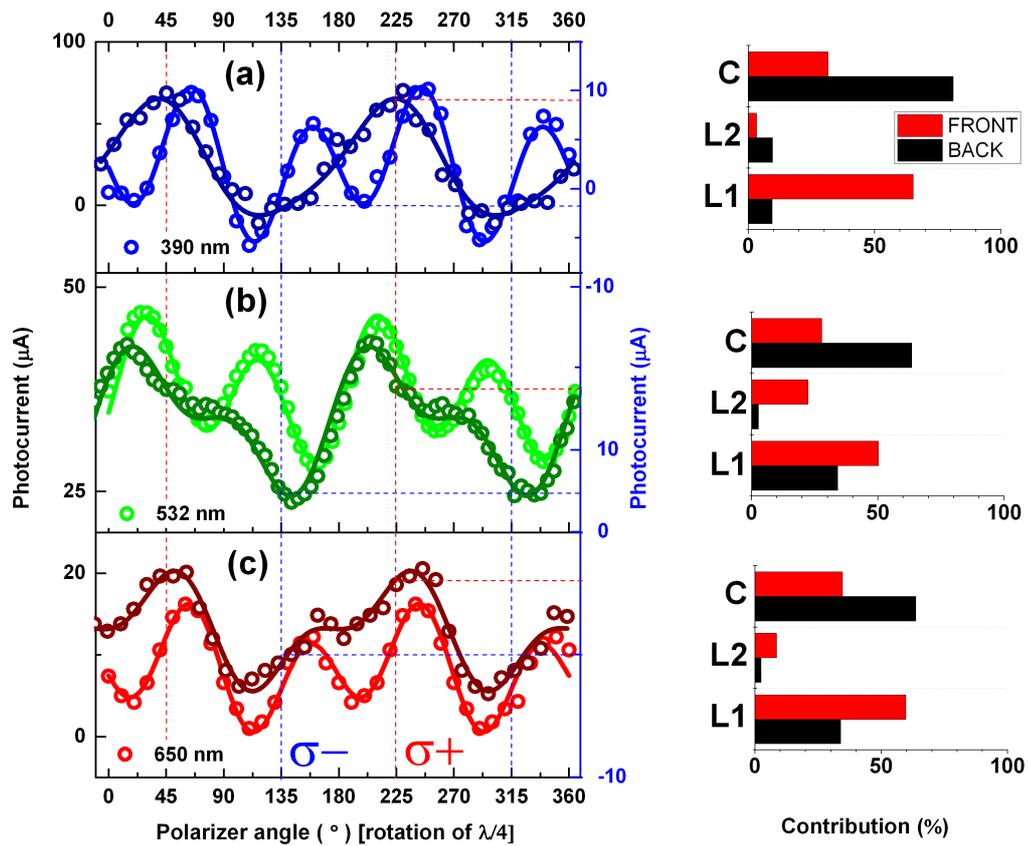

Figure 2 | Wavelength dependent contributions of circular and linear polarization sensitive photocurrents. Open circles represent the experimental data obtained from the MAPI sample excited with (a) a Second harmonic femtosecond Ti:Sa laser (390 nm) and CW laser diodes operating at 532 nm (b) and 650 nm (c). The photocurrents demonstrate different behavior when the sample is excited from front (lighter color) and back (darker color) sides. The percentile contributions of different types of photocurrents extracted from fitting are shown on bar plots in the right side of the figure. Coefficients *C, L1, L2* are the fitting parameters from equation (2). Fitting is shown with solid lines.

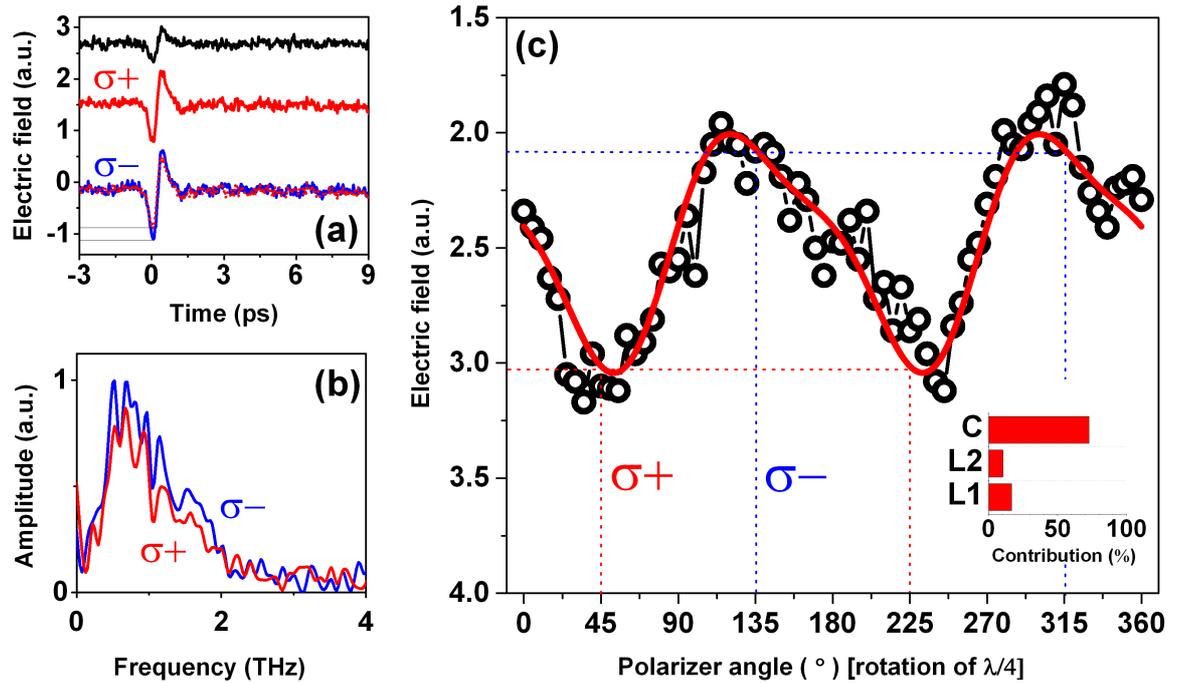

Figure 3 | Polarization dependent coherent terahertz emission from unbiased MAPI at room temperature. Excitation of the MAPI sample with a femtosecond laser pulses leads to efficient emission of THz radiation. The typical transient waveforms of the emitted THz radiation detected with a GaP EO crystal at the far field under oblique excitation with linear (black curve), right circular (red curve), and left circular (blue curve) excitation beam polarization is shown in (a). Typical THz frequency spectra obtained by Fourier transform of the transient waveforms for the case of right circular (red curve) and left circular (blue curve) polarizations are shown in (b). (c) Photon-polarization dependence of the emitted THz field amplitude and fitting with equation (2) are shown with open circles and solid line correspondingly. The fitting results are shown on the bar plot in the right part of (c).